\documentclass{cimento}

\usepackage{cite}
\def\lambdabar {\mathchar'26\mkern-10mu\lambda}

\title{The relativistic two-dimensional harmonic oscillator}

\author{S.M.~Nagiyev\from{ins:p}\ETC,
E.I.~Jafarov\from{ins:p}\thanks{E-mail: ejafarov@physics.ab.az}
        \atque
M.Y.~Efendiyev\from{ins:c}}
\instlist{\inst{ins:p} Institute of Physics, Azerbaijan National Academy of Sciences, Javid av. 33, AZ1143,
Baku, Azerbaijan
  \inst{ins:c} Azerbaijan Cooperation University, Narimanov av. 86, AZ1106, 
Baku, Azerbaijan}
\PACSes{\PACSit{03.65.Db}{Functional analytical methods}
\PACSit{03.65.Pm}{Relativistic wave equations}\PACSit{02.70.Bf}{Finite-difference methods}}

\begin{document}

\maketitle

\begin{abstract}
The two-dimensional relativistic configurational $\vec r$-space is proposed and the exactly solvable finite-difference model of the harmonic oscillator in this space is constructed. The wave functions of the stationary states and the corresponding energy spectrum are found for the model under consideration. It is shown that, they have correct non-relativistic limit. Explicit form of the plane momentum operator is determined.
\end{abstract}


\section{Introduction}

The harmonic oscillator is one of the important problems of the non-relativistic quantum physics. It is one of few problems of the quantum mechanics having an exact solutions. The theory of the harmonic oscillator has various applications in the atomic and molecular as well as nuclear and particle physics~\cite{moshinsky}. 

Usually, analytical solutions of the one- and three-dimensional non-relativistic quantum oscillator models are well known due to their enormous importance for explanation of the various quantum physics phenomena. The solution of the stationary Schr\"odinger equation for the one-dimensional non-relativistic oscillator is simple leading to the wave functions expressed through the Hermite polynomials, whereas the similar three-dimensional solution leads to the radial wave function that is expressed through the generalized Laguerre polynomials. One need to note here the two-dimensional model of the non-relativistic quantum harmonic oscillator, that also has an explicit analytical solution for the stationary states similar to the three-dimensional model~\cite{flugge} and this model has a lot of important applications, too. For example, in~\cite{zhu}, the two-dimensional model of the harmonic oscillator is used to study the Aharonov-Bohm effect. This oscillator model is also applied to describe a nonlinear amplifier, exemplified by a two-junction superconducting quantum interference device in the presence of thermal noise~\cite{acebron} and its perturbed model is employed to obtain the Rayleigh-Schr\"odinger energy perturbation series~\cite{killingbeck}. One can note a recent interesting work~\cite{carinena}, that considers different two-dimensional oscillators, namely a nonlinear Lachmanan-Mathews oscillator, a nonlinear oscillator related to the Riccati equation and the standard harmonic oscillator on constant curvature spaces as nonlinear deformations of the linear harmonic oscillator as well as another quite interesting work~\cite{jafarpour} uses a variational method with a normal ordering technique
to express the Hamiltonian of two-dimensional anharmonic oscillators, in terms of Wick's ordered-product of the functions of the generalized creation and annihilation operators in order to find the eigenvalues of the 2D anharmonic Hamiltonian.

However, almost all of these studies has been done in the framework of the non-relativistic quantum approach. To consider mentioned above the two-dimensional problem in the framework of the relativistic quantum approach, one of the ways is to solve Dirac or Klein-Gordon equations analytically. An approach to solve the Dirac equation for quantum harmonic oscillator, that proposed in~\cite{moshinsky2} allows to consider Dirac oscillator in two-dimensions, too~\cite{villalba}, where the problem has explicit solutions under requirement that the potential should be introduced as a pseudovectorial term shifting the momentum operator. However, this is not only the way to construct the two-dimensional relativistic oscillator model, and another elegant relativistic approach is the relativistic quantum mechanics based on the finite-difference equations.

The purpose of this paper is to construct the relativistic  model of the two-dimensional harmonic oscillator~\cite{flugge}. Our construction is based on the relativistic finite-difference quantum mechanics, developed in~\cite{donkov,atakishiyev1,atakishiyev-protvino,atakishiyev2,kadyshevsky1,kadyshevsky2a,freeman,kadyshevsky3,amirkhanov,atakishiyev3,kagramanov1}. Here, the relativistic configuration space is the key conception~\cite{kadyshevsky1}. Note that, the models of the linear, three-dimensional and $N$-dimensional harmonic oscillators already have been studied in the framework of the relativistic finite-difference quantum mechanics~\cite{donkov,atakishiyev1,atakishiyev-protvino,atakishiyev2,atakishiyev-preprint}.

The structure of the paper is as follows: We consider two-dimensional relativistic configurational $\vec r\left( {x,y} \right)$-space in the second section. The model of the  harmonic oscillator in this space is proposed in the third section.

\section{Two-dimensional relativistic configurational $\vec r\left( {x,y} \right)$-space}

Two-dimensional relativistic equation for the wave function in the momentum $\vec p\left( {p_x,p_y} \right)$-representation in the case of the equal masses has the following form~\cite{atakishiyev1,kadyshevsky1}:

\begin{equation}
\label{2.1}
\left( {E_p  - E} \right)\psi \left( {\vec p} \right) = \frac{1}{{\left( {2\pi } \right)^2 }}\int {V\left( {\vec p,\vec k;E} \right)\psi \left( {\vec k} \right)d\Omega _k } ,
\end{equation}
where $d\Omega _k  = d\vec k/\sqrt {1 + \vec k^2 /m^2 c^2 } $ is an invariant volume element of the integration in the $\vec p$-representation, $E_p  = c\sqrt {\vec p^2  + m^2 c^2 } $ is the relativistic energy and $V\left( {\vec p,\vec k;E} \right)$ is the interaction potential.

We carry out the integration in the (\ref{2.1}) over the mass shell of the particle with mass m, i.e. over the upper sheet of the hyperboloid $p_0 ^2  - \vec p^2  = m^2 c^2 $ (two-dimensional Lobachevsky space).

Transition to relativistic two-dimensional configuration $\vec r$-space

\begin{equation}
\label{2.2}
\psi \left( \vec r\right) =\frac 1{ 2\pi \hbar}\int \xi \left( \vec p,\vec r\right) \psi \left( \vec p\right) d\Omega _p
\end{equation}
is performed by the use of expansion on the matrix elements of the representations of the two-dimensional Lobachevsky space:

\begin{equation}
\label{2.3}
\xi \left( \vec p,\vec r\right) =\left( \frac{p_0-\vec p\vec n}{mc}\right) ^{-\frac 1 2 -ir/\lambdabar } \; ,
\end{equation}
where $\vec r=r\vec n,\quad 0\leq r<\infty $, $\vec n=\left(\cos \varphi , \sin \varphi ,\right)$ , $p_0=E_p/c$ and $\lambdabar=\hbar/mc$ is the Compton wavelength of the particle.

As a result of this transition, the eq. (\ref{2.1}) in the case of the local potentiald of the interaction has following finite-difference form:

\begin{equation}
\label{2.4}
\left[ H_0+V\left( \vec r \right) \right] \psi (\vec r)=E\psi (\vec r)
\end{equation}
with the free Hamiltonian $H_0$ of the form:

\begin{equation}
\label{2.5}
H_0  = mc^2 \left[ {\cosh \left( {i\lambdabar \partial _r } \right) + \frac{{i\lambdabar }}{{2r}}\sinh \left( {i\lambdabar \partial _r } \right) - \frac{{\lambdabar ^2 }}{{r\left( {2r + i\lambdabar } \right)}}\partial _\varphi  ^2 e^{i\lambdabar \partial _r } } \right].
\end{equation}

Here, the  momentum operator of the free particle has following explicit determination:

\begin{equation}
\label{2.6}
\hat {\vec p} = mc \cdot \vec n\left( {\frac{{H_0 }}{{mc^2 }} - e^{i\lambdabar \partial _r } } \right) - \vec m\frac{\hbar }{{r + i\lambdabar /2}}e^{i\lambdabar \partial _r } ,
\end{equation}
that is expressed by emploing the two-dimensional vector $\vec m$ of the following components:

\[
\vec m = i\left( { - \sin \varphi  \cdot \partial _\varphi  ,\cos \varphi  \cdot \partial _\varphi  } \right).
\]

Some useful relations for the operators $\vec n$, $\vec m$ and $\hat L=-i\partial _\varphi$ are presented in the Appendix. The components of the $\hat {\vec p}$ momentum vector and free Hamiltonian $H_0$ commutate as follows:

\[
\left[ {\hat p_i ,\hat p_j } \right] = \left[ {\hat {\vec p}_i ,H_0 } \right] = 0,\quad i,j = x,y.
\]

The function (\ref{2.3}) being the relativistic 'plane wave' has the property to be common eigenfunction of the operators $H_0$ and $\hat {\vec p}$, i.e.

\begin{equation}
\label{2.7}
H_0  \cdot \xi \left( {\vec p,\vec r} \right) = E_p  \cdot \xi \left( {\vec p,\vec r} \right),\quad \hat {\vec p} \cdot \xi \left( {\vec p,\vec r} \right) = \vec p \cdot \xi \left( {\vec p,\vec r} \right).
\end{equation}

The relativistic plane waves (\ref{2.3}) form orthogonal and complete system of the functions:

\begin{eqnarray}
 \frac{1}{{\left( {2\pi \hbar } \right)^2 }}\int {\xi ^* \left( {\vec p,\vec r} \right) \cdot \xi \left( {\vec p,\vec r'} \right)d\Omega _p }  = w^{ - 1} \left( r \right) \cdot \delta \left( {\vec r - \vec r'} \right), \nonumber \\ 
 \label{2.8} \\ 
 \frac{1}{{\left( {2\pi \hbar } \right)^2 }}\int {\xi ^* \left( {\vec p,\vec r} \right) \cdot \xi \left( {\vec p',\vec r} \right)w\left( r \right)d\vec r}  = \delta \left( {\vec p\left(  -  \right)\vec p'} \right),\nonumber
\end{eqnarray}
where, $w\left( r \right) = \frac{\lambdabar }{r}\left| {\left( { - r/\lambdabar } \right)^{\left( {1/2} \right)} } \right|^2 $ is the weight function~\cite{atakishiyev-preprint}, $\delta \left( {\vec p\left(  -  \right)\vec p'} \right) = \frac{{p_0 }}{{mc}}\delta \left( {\vec p - \vec p'} \right)$, and $r^{\left( \delta \right)}$ is the generalized degree~\cite{kadyshevsky1} having the form:

\[
\left( {\frac{r}{\lambdabar }} \right)^{\left( \delta  \right)}  = i^\delta  \frac{{\Gamma \left( { - ir/\lambdabar  + \delta } \right)}}{{\Gamma \left( { - ir/\lambdabar } \right)}}.
\]

In the non-relativistic limit we have the following reducing relations:

\begin{eqnarray}
 \mathop {\lim }\limits_{c \to \infty } \left( {H_0  - mc^2 } \right) = H_o ^{NR}  =  - \frac{{\hbar ^2 }}{{2m}}\left( {\partial _r ^2  + \frac{1}{r}\partial _r  - \frac{1}{{r^2 }}\partial _\varphi  ^2 } \right), \nonumber \\ 
\label{2.9}
 \mathop {\lim }\limits_{c \to \infty } \hat {\vec p} = \hat {\vec p_{NR}}  =  - \hbar \left( {i\vec n\partial _r  + \vec m\frac{1}{r}} \right), \\ 
 \mathop {\lim }\limits_{c \to \infty } \xi \left( {\vec p,\vec r} \right) = e^{i\vec p\vec r} . \nonumber
\end{eqnarray}

\section{The relativistic finite-difference model of the two-dimensional harmonic oscillator}

We consider the model of the relativistic two-dimensional oscillator that corresponds to the following potential of the interaction:

\begin{equation}
\label{3.1}
V\left( r \right) = \frac{{m\omega ^2 }}{2} \cdot \frac{{r + i\lambdabar }}{{r + i\lambdabar /2}}\left[ {r\left( {r + i\lambdabar } \right) - \lambdabar ^2 b \cdot \partial _\varphi  ^2 } \right]e^{i\lambdabar \partial _r } .
\end{equation}

It is clear that in the non-relativistic limit we will have:

\[
\mathop {\lim }\limits_{c \to \infty } V\left( r \right) = \frac{1}{2}m\omega ^2 r^2 .
\]

Due to commutativity of the operators $H=H_0+V\left( r \right)$ and $\hat L =  - i\partial _\varphi  $, eigenfunctions of the Hamiltonian $H$ depend on the angle $\varphi$ by standard manner:

\begin{equation}
\label{3.2}
\psi \left( {\vec r} \right) = \left[ {\left( { - r/\lambdabar } \right)^{\left( {1/2} \right)} } \right]^{ - 1} R_{\left| m \right|} \left( r \right)\frac{{e^{im\varphi } }}{{\sqrt {2\pi } }},\quad m = 0, \pm 1, \pm 2, \ldots \;.
\end{equation}

Therefore, considered two-dimensional problem will be reduced to the linear one, i.e. we have to find the eigenfunctions and eigenvalues of the radial part of the Hamiltonian

\begin{equation}
\label{3.3}
\tilde H \cdot R_{\left| m \right|} \left( r \right) = E_{\left| m \right|}  \cdot R_{\left| m \right|} \left( r \right)
\end{equation}
with the boundary conditions $R_{\left| m \right|} \left( 0 \right) = R_{\left| m \right|} \left( \infty  \right) = 0$. Here $\tilde H$ has following explicit finite-difference form:

\begin{equation}
\label{3.4}
\quad \tilde H = mc^2 \left\{ {\cosh \left( {i\lambdabar \partial _r } \right) + \frac{{\lambdabar ^2 \left( {m^2 - 1/4} \right)}}{{2r\left( {r + i\lambdabar } \right)}}e^{i\lambdabar \partial _r }  + \frac{1}{2}m\omega ^2 \left[ {r\left( {r + i\lambdabar } \right) + \lambdabar ^2 bm^2 } \right]e^{i\lambdabar \partial _r } } \right\}.
\end{equation}

In terms of the dimensionless variable $\rho = r/ \lambdabar$ and parameter $\omega _0  = \hbar \omega /mc^2 $, the eq. (\ref{3.3}) will have the following form:

\begin{equation}
\label{3.5}
\left[ {\cosh \left( {i\partial _\rho  } \right) + \frac{a}{{2\rho ^{\left( 2 \right)} }}e^{i\partial _\rho  }  + \frac{1}{2}\omega _0 ^2 \left( {\rho ^{\left( 2 \right)}  + \gamma } \right)e^{i\partial _\rho  } } \right]R_{\left| m \right|} \left( \rho  \right) = \frac{{E_{\left| m \right|} }}{{mc^2 }}R_{\left| m \right|} \left( \rho  \right),
\end{equation}
with $a=m^2 - 1/4$ and $\gamma=bm^2$.

According to the works~\cite{donkov,atakishiyev-protvino,nagiyev2}, we extract multipliers $\left( { - \rho } \right)^{\left( \alpha \right)} $ and $M_\nu  \left( \rho  \right) = \omega _0 ^{i\rho } \Gamma \left( {i\rho  + \nu } \right)$, which determine the behaviour of the radial wave function $R_{\left| m \right|} \left( \rho  \right)$ at corresponding points $\rho=0$ and $\rho= \infty$, and therefore we look for solution of the {(\ref{3.5}) as:

\begin{equation}
\label{3.6}
R_{\left| m \right|} \left( \rho  \right) = C_{\left| m \right|}  \cdot \left( { - \rho } \right)^{\left( \alpha  \right)}  \cdot M_\nu  \left( \rho  \right) \cdot \Omega \left( \rho  \right),
\end{equation}
where the constants $\alpha$ and $\nu$ equal to:

\begin{eqnarray}
\label{3.7}
 \alpha  = \frac{1}{2} + \frac{1}{2}\sqrt {1 + 2\left( {\gamma  + \frac{1}{{\omega _0 ^2 }} - \sqrt {\left( {\gamma  + \frac{1}{{\omega _0 ^2 }}} \right)^2  - \frac{{4a}}{{\omega _0 ^2 }}} } \right)} , \\ 
\label{3.8}
 \nu  = \frac{1}{2} + \frac{1}{2}\sqrt {1 + 2\left( {\gamma  + \frac{1}{{\omega _0 ^2 }} + \sqrt {\left( {\gamma  + \frac{1}{{\omega _0 ^2 }}} \right)^2  - \frac{{4a}}{{\omega _0 ^2 }}} } \right)} . 
\end{eqnarray}

Then, the function $\Omega \left( \rho  \right)$ will satisfy the following finite-difference equation:

\begin{equation}
\label{3.9}
\left[ {\left( {\alpha  + i\rho } \right)\left( {\nu  + i\rho } \right)e^{ - i\partial _\rho  }  - \left( {\alpha  - i\rho } \right)\left( {\nu  - i\rho } \right)e^{i\partial _\rho  } } \right]\Omega \left( \rho  \right) = 2i\frac{{E_{\left| m \right|} }}{{\hbar \omega }}\Omega \left( \rho  \right).
\end{equation}

Polynomial solutions of the eq. (\ref{3.9}), corresponding to the energy spectrum values

\begin{equation}
\label{3.10}
E_{\left| m \right|}  \equiv E_{n\left| m \right|}  = \hbar \omega \left( {2n + \alpha  + \nu } \right),\quad n = 0,1,2, \ldots 
\end{equation}
are expressed through the continuous dual Hahn polynomials~\cite{nagiyev2}:

\begin{equation}
\label{3.11}
\Omega \left( \rho  \right) \equiv \Omega _n \left( \rho  \right) = S_n \left( {\rho ^2 ;\alpha ,\nu ,1/2} \right).
\end{equation}

The condition (\ref{3.10}) gives the quantization rule for the energy level of the two-dimensional harmonic oscillator (\ref{3.1}) and leads to the following form of the radial part of the wave function:

\begin{eqnarray}
\label{3.12}
 R_{n\left| m \right|} \left( \rho  \right) = C_{n\left| m \right|}  \cdot \left( { - \rho } \right)^{\left( \alpha  \right)}  \cdot M_\nu  \left( \rho  \right) \cdot S_n \left( {\rho ^2 ;\alpha ,\nu ,1/2} \right), \\ 
 C_{n\left| m \right|}  = \sqrt {\frac{2}{{n! \cdot \Gamma \left( {n + \alpha  + \nu } \right) \cdot \Gamma \left( {n + \alpha  + 1/2} \right) \cdot \Gamma \left( {n + \nu  + 1/2} \right)}}} . \nonumber
\end{eqnarray}

The functions $R_{n\left| m \right|} \left( \rho  \right)$ are orthogonal and normalized by following manner:

\begin{equation}
\label{3.13}
\int\limits_0^\infty  {R_{n\left| m \right|} \left( \rho  \right) \cdot R_{n'\left| m \right|} \left( \rho  \right) \cdot d\rho }  = \delta _{nn'} .
\end{equation}

Therefore, from (\ref{3.13}) we obtain following orthonormalization condition for the full wave function (\ref{3.2}):

\begin{equation}
\label{3.14}
\int {\psi _{n\left| m \right|} ^* \left( {\vec r} \right) \cdot \psi _{n'\left| {m'} \right|} \left( {\vec r} \right) \cdot w\left( r \right)d\vec r}  = \delta _{nn'} \delta _{mm'} .
\end{equation}

In the non-relativistic limit, we find that

\begin{eqnarray}
 \mathop {\lim }\limits_{c \to \infty } \alpha  = \frac{1}{2} + \frac{1}{2}\sqrt {1 + na}  = \frac{1}{2} + \left| m \right|, \nonumber \\ 
 \mathop {\lim }\limits_{c \to \infty } \left( {\nu  - \frac{1}{{\omega _0 }}} \right) = \frac{1}{2}, \nonumber \\ 
\label{3.15}  \\
 \mathop {\lim }\limits_{c \to \infty } w\left( r \right) = 1, \nonumber \\ 
 \mathop {\lim }\limits_{c \to \infty } \frac{{\omega _0 ^n }}{{n!}}S_n \left( {\xi ^2 /\omega _0 ;\alpha ,\nu ,1/2} \right) = L_n^{\left| m \right|} \left( {\xi ^2 } \right). \nonumber
\end{eqnarray}

Here, $\rho ^2  = \xi ^2 /\omega _0 $, $\xi  = r\sqrt {\frac{{m\omega }}{\hbar }}$ and  $L_n^\mu \left( x \right)$ are the generalized Laguerre polynomials:

\[
L_n^\mu  \left( x \right) = \frac{{\left( {\mu  + 1} \right)_n }}{{n!}} \cdot _1 F_1 \left( { - n;\mu  + 1,x} \right).
\]

Consequently, in the non-relativistic limit $c \to \infty$, wave functions (\ref{3.2}) and energy spectrum (\ref{3.10}) coincide with the corresponding wave functions and energy spectrum of the non-relativistic two-dimensional harmonic oscillator, i.e.

\begin{eqnarray}
 \mathop {\lim }\limits_{c \to \infty } \psi _{n\left| m \right|} \left( {\vec r} \right) = \psi _{n\left| m \right|} ^{NR} \left( {\vec r} \right) = C_{n\left| m \right|} ^{NR}  \cdot \xi ^{\left| m \right|} e^{ - \frac{1}{2}\xi ^2 }  \cdot L_n^{\left| m \right|} \left( {\xi ^2 } \right)\frac{{e^{im\varphi } }}{{\sqrt {2\pi } }}, \nonumber \\ 
\label{3.16}  \\ 
 \mathop {\lim }\limits_{c \to \infty } \left( {E_{n\left| m \right|}  - mc^2 } \right) = E_{n\left| m \right|} ^{NR}  = \hbar \omega \left( {2n + \left| m \right| + 1} \right). \nonumber
\end{eqnarray}

The wave functions $\psi _{n\left| m \right|} ^{NR} \left( {\vec r} \right)$ are othonormalized as follows:

\begin{equation}
\label{3.17}
\int {\left( {\psi _{n\left| m \right|} ^{NR} \left( {\vec r} \right)} \right)^*  \cdot \psi _{n'\left| {m'} \right|} ^{NR} \left( {\vec r} \right)d\vec r}  = \delta _{nn'} \delta _{mm'} .
\end{equation}

From (\ref{3.17}) we can see that normalization constant has following form:

\[
C_{n\left| m \right|} ^{NR}  = \sqrt {\frac{{2m\omega n!}}{{\hbar \Gamma \left( {n + \left| m \right| + 1} \right)}}} .
\]

\appendix
\section{}

The operators $\vec n$, $\vec m$ and $\hat L = - i \partial_\varphi$ satisfy following relations:

\[
\left[ {n_x ,\hat L} \right] =  - in_y ,\quad \left[ {n_y ,\hat L} \right] = in_x ,\quad \left[ {m_x ,\hat L} \right] =  - im_y ,\quad \left[ {m_y ,\hat L} \right] = im_x ,
\]
\[
\left[ {n_i ,\hat L^2 } \right] =  - \left( {n_i  + 2i \cdot m_i } \right),\quad \left[ {m_i ,\hat L^2 } \right] =  - \left( {m_i  + 2i \cdot n_i \hat L^2 } \right),
\]
\[
\left[ {n_y ,m_x } \right] = \left[ {n_x ,m_y } \right] = in_x n_y ,
\]
\[
\left[ {m_x ,m_y } \right] =  - i\hat L,\quad \left[ {m_x ,n_x } \right] = in_y ^2 ,\quad \left[ {m_y ,n_y } \right] = in_y ^2 .
\]

Their scalar products are:

\[
\vec n \cdot \vec m = 0,\quad \vec m \cdot \vec n = i,\quad \vec m^2  = \hat L^2 .
\]

Under the Hermitian conjugation with respect to the scalar product $\int {\phi ^*  \cdot \psi d\vec r} $, we have $n_i ^ +   = n_i $, $m_i ^ +   =i n_i -m_i$, $i=x,y$.

\end{document}